%% file: MBT.tex
\documentclass[submission,copyright,creativecommons]{eptcs}
\providecommand{\event}{MBT 2013}
 \let\originalvec=\vec
\let\vec=\originalvec

 \usepackage{amsmath}
\usepackage{amssymb}
 \usepackage{amsthm}
 \usepackage{multirow}
 \usepackage{rotating}

\usepackage{algorithmic}
\usepackage{algorithm2e}
\usepackage{latexsym}
\usepackage{graphicx}
\usepackage{listings}
\usepackage{framed}
\usepackage{times}
\usepackage{pdfsync}
\usepackage{url}
\input{macros.tex}

\title{Testing Java implementations of algebraic specifications}
\author{Isabel Nunes
\institute{\small Faculty of Sciences, University of Lisbon \\
Lisboa, Portugal }
\email{in@di.fc.ul.pt}
\and
Filipe Lu\'\i s
\institute{\small Faculty of Sciences, University of Lisbon \\
Lisboa, Portugal }
\email{\quad fluis@di.fc.ul.pt}
}
\def\titlerunning{Testing Java implementations of algebraic specifications}
\def\authorrunning{Isabel Nunes \& Filipe Lu\'\i s}
\begin{document}
\maketitle

\begin{abstract}

In this paper we focus on exploiting a specification and the
structures that satisfy it, to obtain a means of comparing 
implemented and expected behaviours and find the origin of faults in
implementations. We present an approach to the creation of tests that
are based on those specification-compliant structures, and to the
interpretation of those tests' results leading to the discovery of
the method responsible for an eventual test failure. Results of
comparative experiments with a tool implementing this approach are presented.

\end{abstract}


\section{Introduction}

The development and verification of software programs against specifications of desired properties is growing weight among software engineering methods and tools for promoting software reliability. In particular, finding the software element containing a given fault is highly desirable and several approaches exist that tackle this issue, that can be quite different in the way they approach the problem.

ConGu~\cite{Nunes2009,Nunes2006} is both an approach and a tool for the runtime verification of Java implementations of algebraic specifications. It verifies that implementations conform to specifications by monitoring method executions in order to find any violation of automatically generated pre and post-conditions. 

The ConGu tool~\cite{Crispim2010} picks a module of axiomatic specifications, together with a Java implementation and a refinement that maps specifications to Java types, and responds to an erroneous implementation by outputing the specification constraint that was violated; this is often insufficient to find the faulty method, because all methods involved in the violated constraint become equally suspect.

A ConGu companion tool -- the GenT tool~\cite{Andrade2012,Andrade2011} -- generates JUnit test cases from ConGu specifications.
Generating test cases that are known to be comprehensive, i.e. that cover all constraints of the specification, as GenT does, is a very important activity, because the confidence we may gain on the correction of the software we use greatly depends on it. But, in order for these tests to be of effective use, we should be able to use their results to localize the faulty components. Here again, executing the JUnit tests generated by GenT fails to give the programmer clear hints about the faulty method -- all methods used in failed tests are suspect. The ideal result of a test suite execution would be the exact localization of the fault.

In this paper we enrich ConGu, by giving it the capability of locating the methods that are responsible for detected faults. We present a technique that builds upon structures satisfying the specification to obtain a means to observe the implemented behaviour against the intended one, and to locate faulty methods in implementations. Unlike several existing approaches, ours does not inspect the executed code; instead, it exploits the specification and conforming structures in order to be able to interpret some failures and discover their origin. 

A tool was built -- the Flasji tool -- that implements the presented technique, and a comparative experiment was undertaken to evaluate its results. A summary of these results, which were very encouraging, is presented in this paper.

The unit of fault Flasji is able to detect is the method, leaving to the programmer the task of identifying the exact instruction within it that is faulty. If more than one fault exists, the repeated application of the process, together with the correction of the identified faulty method, should be adopted.
In what concerns integration testing strategy, Flasji applies an incremental one in the sense that the Java types implementing the specification are not tested all together; instead, each one is tested conditionally, presuming all others from which it depends are correctly implemented. This incremental integration is possible since the overall specification is given as a structured collection of individual specifications (a ConGu module), whose structure is matched by the structure of the Java collection of classes that implements it.

The remainder of the paper is organized as follows: section~\ref{sec:prelim} introduces the ConGu specification language through an example that will be used throughout the paper, and gives an overview of the Flasji approach; section~\ref{sec:abstractvsconcrete} details every Flasji step, from picking a specification module and corresponding implementation, to the identification of the faulty method, explaining the several items Flasji produces; an evaluation experiment of the Flasji tool is presented in section~\ref{sec:evaluation} where results are compared with the ones obtained using two other tools; in section~\ref{sec:rw}  we  focus our discussion on relevant aspects related to the work presented in this paper; finally, section~\ref{sec:conclusoes} concludes.


\section{Approach Overview}
\label{sec:prelim}

In this section we give a general
overview of the Flasji approach. An
example is introduced that will be used throughout the paper.

\begin{figure*}
\centering
\includegraphics[width=1\textwidth]{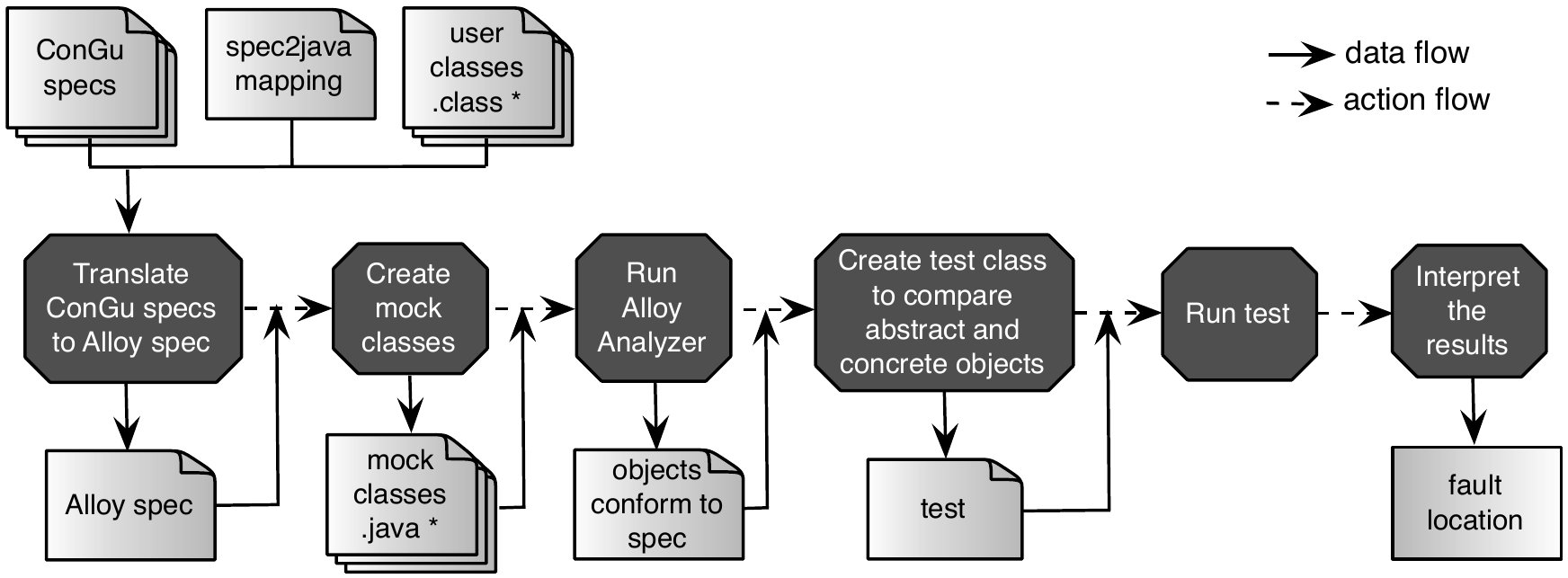}
\caption{Overview of the Flasji approach.}
\label{fig:overview}
\end{figure*} 

\subsection{The approach in a nutshell}
\label{sec:overview}

As illustrated in figure~\ref{fig:overview}, the Flasji approach integrates a
series of steps from an initial input comprising a module of ConGu specifications and
corresponding Java implementations (together with a mapping defining
correspondences between the two), to a final output comprising the
method identified as the one containing the fault, whenever possible, and a
list of other methods suspect of being faulty. The whole process leading
from the initial input to the final output is automated, without any
further user intervention.

The main strategy underlying the Flasji process is the comparison between what we
call ``abstract'' and ``concrete'' objects; the former are objects
that are well-behaved in the sense that they conform to the specification,
while the latter are objects that behave according to the classes implementing the
specification, which we want to investigate for faults.

We capitalize on the Alloy Analyzer~\cite{Alloy}
tool which is capable of finding structures that satisfy a collection
of constraints -- a specification. The
specifications this tool works with are written in the
Alloy~\cite{Jackson2012} language.

Flasji begins by translating the ConGu specification module into an Alloy
specification, in order to be able to, in a posterior phase, generate structures
satisfying it.  It then creates Java classes whose instances will
represent objects satisfying the specification (the ``abstract''
objects) -- these classes are called
``mock'' classes;
in order for ``abstract'' objects to represent structures that satisfy the
specification, they are given the ability of storing and
retrieving the results of applying each and every operation of the
specification, as will be seen later.

A third step feeds the Alloy Analyzer
tool with the specification, asking the tool for a collection of structures
satisfying the specification. This collection will
be used in the next step to define the abstract objects, which will present the expected, correct, behaviours. 
 
In a fourth step, a test class is created that contains instructions to instantiate both
the mock classes and the implementation classes given as input, and to
compare
the behaviour of the ``concrete'' objects against the ``abstract'' ones, in order to identify the
faulty method. Flasji then executes this test class and interprets its results
to obtain the faulty method. 

Remember that all these steps are automatically processed, thus
transparent to the Flasji user. Section~\ref{sec:abstractvsconcrete} describes them in detail.

\subsection{A specification and corresponding implementation}
\label{sec:runningExample}
Simple sorts, sub-sorts and parameterized sorts can be specified with the ConGu specification language, and mappings between those sorts and Java types, and between those sorts' operations and Java methods, can be defined using the ConGu refinement language.

\begin{figure*}
\centering
\includegraphics[width=0.8\textwidth]{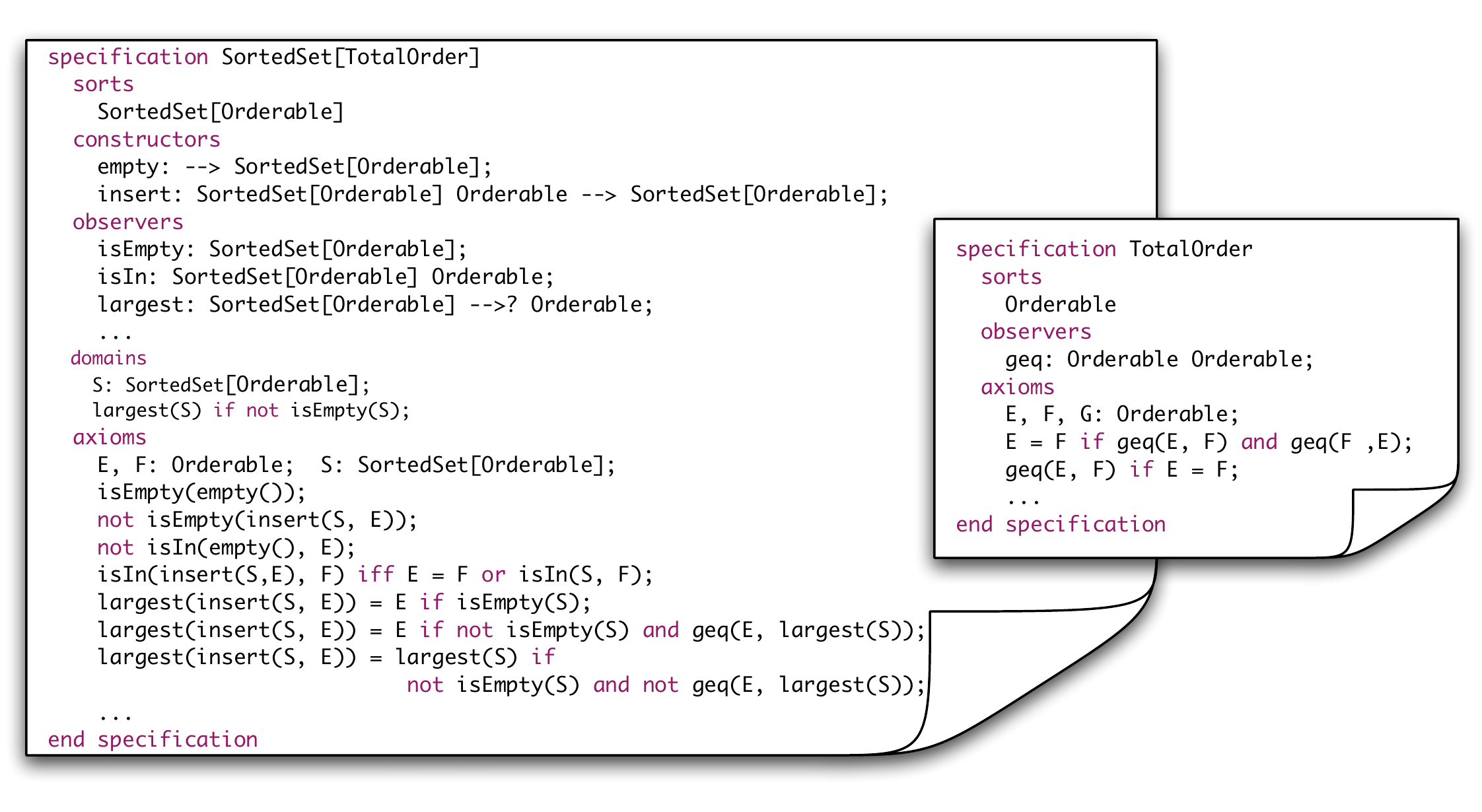}
\caption{Parts of the ConGu specifications for the \congutext{SortedSet}
parameterized data type and its parameter.}
\label{fig:SortedSet}
\end{figure*}

We present a classical yet rich example, of a ConGu specification of the
\congutext{SortedSet} parameterized data type, representing a set of
\congutext{Orderable} elements, together with the
specification for its parameter (figure~\ref{fig:SortedSet}). 

In a specification, we define \textit{constructors}, \textit{observers} and other
operations,
where constructors compose the minimal set of operations that allow to
build all instances of the sort, observers can be used to analyse those instances, and the other operations are usually comparison operations or operations derived from the others; depending on whether they have an
argument of the sort under specification (self argument) or not, constructors are classified as
\textit{transformers} or \textit{creators} (transformers are also
referred to as \textit{non-creator constructors}). 

All operations that are not constructors must have a self argument. Any function can be partial
(denoted by $\rightarrow ?$), in which case a \textit{domains} section
specifies the conditions that restrict their domain. \textit{Axioms} define
every value of the type through the application of observers to
constructors
-- see, e.g., the axiom
\congutext{isEmpty(empty());} that specifies that the result of applying
the observer operation \congutext{isEmpty} to a \congutext{SortedSet}
instance obtained with the creator constructor \congutext{empty} is
\congutext{true}, and the axiom \congutext{not isEmpty(insert(S, E));}
saying that the result of applying \congutext{isEmpty} to any \congutext{SortedSet}
instance to which the transformer constructor \congutext{insert} has
been applied is \congutext{false}.

\begin{figure*}
\centering
\includegraphics[width=0.7\textwidth]{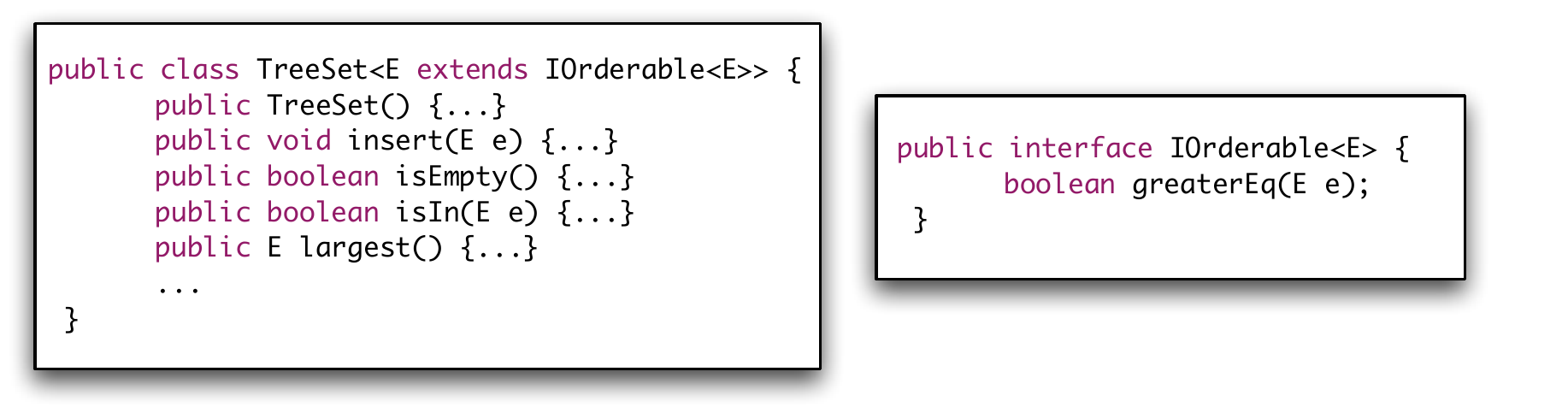}
\caption{Excerpt from a Java implementation of the ConGu specification
for \congutext{SortedSet}.}
\label{fig:Implementation}
\end{figure*}

Generic Java class \javatext{TreeSet} in figure~\ref{fig:Implementation}
represents
a Java implementation of the
\congutext{SortedSet} parameterized data type; in the same figure, interface
\javatext{IOrderable} represents a Java type restraining the
\javatext{TreeSet} parameter. We want to
investigate the \javatext{TreeSet} class for faults, independent of
any specific
implementation of its parameter type. 

\begin{figure*}
\centering
\includegraphics[width=0.7\textwidth]{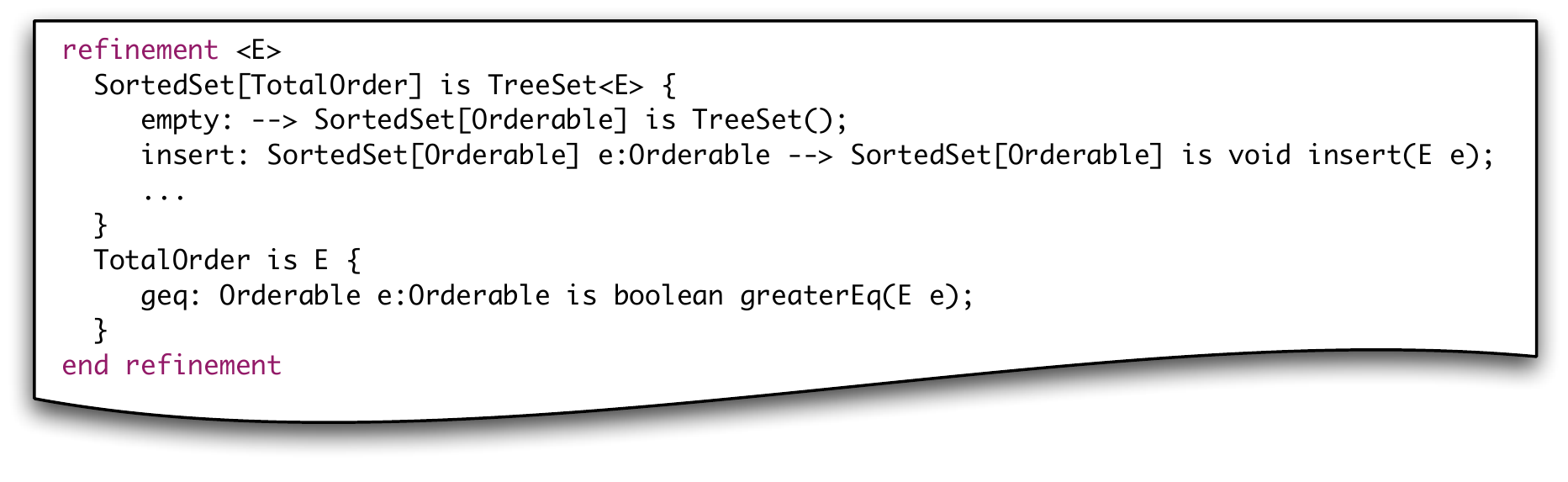}
\caption{Refinement mapping from ConGu specifications to Java types.}
\label{fig:Refinement}
\end{figure*}

The correspondence between ConGu and Java types must be defined
in order for implementations to be checked. This correspondence is described in
terms of \textit{refinement mappings}; figure~\ref{fig:Refinement} shows a
refinement mapping from the specifications \congutext{SortedSet} and \congutext{TotalOrder} 
(figure~\ref{fig:SortedSet}) to Java types \javatext{TreeSet} and
\javatext{IOrderable} (figure~\ref{fig:Implementation}).

These mappings associate ConGu sorts and operations to Java types and
corresponding methods. The \congutext{insert} operation of sort
\congutext{SortedSet} is mapped to the \javatext{TreeSet} class method
with the same name with the signature 
\javatext{void insert(E e)}. Notice that the \congutext{TotalOrder} parameter sort is mapped to a Java
type variable that is used as the parameter of the generic
\javatext{TreeSet} implementation; this specific mapping is
interpreted as constraining any instantiation of the \javatext{TreeSet} parameter to a
Java type \javatext{Some} containing a method with signature
\javatext{boolean greaterEq(Some e)}. 

Detailed information
about the ConGu approach can be found
in~\cite{Nunes2009,Nunes2006}.

\section{Flasji step-by-step}
\label{sec:abstractvsconcrete}

As already said in the previous section, the main goal of the
Flasji approach is to verify whether ``concrete'' Java objects behave
the same as corresponding ``abstract'' ones; the deviations to the
expected behaviour are interpreted in order to find the location of
the faulty method.

Only one of the implementing classes is under verification -- the
\textit{core} type, that is, the one that implements the \textit{core}
sort --, which is the one at the root of the class association
graph (the \javatext{TreeSet} class in the example). Thus, both
``abstract'' and
``concrete'' objects will be created for this type, in order to
compare behaviours. This does not
apply for non-core
types since they are not
under verification. However, as we shall see, ``abstract'' parameter
objects must be created.

Let us now detail the several steps of the Flasji approach.

\subsection{Translating the ConGu specification module}
\label{sec:translate}
Flasji creates an Alloy specification equivalent to the ConGu specification module, in order to be able, ahead in the process, to obtain a
collection of objects that conform to the specification, thus defining
expected, correct behaviour. 

This step capitalizes on already existing
work, referred to in the introduction of this paper, namely
the GenT tool~\cite{Andrade2012,Andrade2011}, of which Flasji uses the
ConGuToAlloy module.

\subsection{Creating mock classes}
\label{sec:mocks}
Flasji creates mock classes for the Java types implementing the
specification core and parameter sorts;
these classes' instances will represent the ``abstract'' objects. In
the running example, two mock classes must be
created, one corresponding to the \javatext{TreeSet} class --
\javatext{TreeSetMock} --, and other corresponding to the
\javatext{IOrderable} interface -- \javatext{OrderableMock}.

This mock class will be used to generate parameter objects that will be inserted not only in ``abstract'' sorted sets (\javatext{TreeSetMock} instances as explained below), but also in ``concrete'' ones (\javatext{TreeSet} instances); the idea, as said before, is to test the implementation of the core signature for any parameter instantiation that correctly implements the \congutext{Orderable} sort.

Each instance of a mock class defines an object conforming to the
specification, including its ``behaviour'', that is, the results of
applying to it all the operations of the type (respecting the
corresponding domain conditions). Since we only compare ``abstract''
and ``concrete'' objects of the core type, fundamental differences
exist between the mock class for this core type and the others. Let us
see first the mock class for the \congutext{Orderable} parameter
of the running example, which is a non-core type.

\begin{lstlisting}[
language=Java, 
breaklines=true,
xleftmargin=5pt,
xrightmargin=0pt,
numbers=left,
numberstyle=\tiny,  
numbersep=5pt,     
captionpos=b, 
caption=Mock class corresponding to the \congutext{Orderable} parameter sort.,
label=fig:OrdMock]
 public class OrderableMock implements IOrderable<OrderableMock>{
   private HashMap<OrderableMock, Boolean> greaterEqResult = new HashMap<OrderableMock, Boolean>();	
   public boolean greaterEq(OrderableMock e) {
     return greaterEqResult.get(e);}
   public void add_greaterEq(OrderableMock e, Boolean result) {
     greaterEqResult.put(e, result);
   }
 }
\end{lstlisting}

For each method \javatext{X} corresponding to a specification operation \congutext{X}, an attribute is defined to keep the information about the results of \javatext{X}, for every combination of the method's parameters (see line 2 for the method \javatext{greaterEq} in interface \javatext{Iorderable}, corresponding to operation \congutext{geq} in sort \congutext{Orderable}); the \javatext{add_X} method ``fills'' that attribute (lines 5 and 6), and the \javatext{X} method retrieves the result for given values of the method's parameters (lines 3 and 4).

The class that represents ``abstract'' objects of the core type (class \javatext{TreeSetMock} in the example) is also generated. The idea here is not to use these ``abstract'' objects on both abstract and concrete contexts, as we do with parameter ones, but to use them to inform us, for every operation, of the results we should expect when applying corresponding methods to corresponding ``concrete'' objects in order to verify whether the latter behave as they should. 

The fundamental difference lies in the information that core type ``abstract'' objects keep for operations whose result is of the core type (\congutext{insert} in the example). Since we want to be able to know whether the ``concrete'' \javatext{TreeSet} object that results from applying the \javatext{insert} method of the \javatext{TreeSet} class to a ``concrete'' object \javatext{concObj} is the correct one, we give the corresponding ``abstract'' object \javatext{absObj} information that allows us to verify it -- we ``feed''  \javatext{absObj} with the ``concrete'' object that should be expected when applying that operation. Ahead in this paper we show how this is achieved; for now, we just present the \javatext{TreeSetMock} mock class in listing~\ref{fig:TreeSetMock}, where attribute and methods in lines 19 to 24 allow ``abstract'' objects to keep and inform about ``concrete'', expected results of applying \javatext{insert} for different values of the method's parameter:

\begin{lstlisting}[
language=Java, 
breaklines=true,
xleftmargin=5pt,
xrightmargin=0pt,
numbers=left,
numberstyle=\tiny,  
numbersep=5pt,     
captionpos=b, 
caption=Mock class corresponding to the \congutext{SortedSet} sort.,
label=fig:TreeSetMock]
public class TreeSetMock <T>{

  // operations whose result is of a non-core type
  private HashMap<T,Boolean> isInResult = new HashMap<T,Boolean>();
  private boolean isEmptyResult;
  private T largestResult;
	
  public boolean isIn (T e){return isInResult.get(e);}	
  public void add_isIn (T e, Boolean result){
    isinResult.put(e, result);}
  public boolean isEmpty(){return isEmptyResult;}	
  public void add_isEmpty(boolean result){
    isEmptyResult = result;}
  public T largest(){return largestResult;}	
  public void add_largest(T result){
    largestResult = result;}

  // operation whose result is of the core type
  private HashMap<T,TreeSet<T>> insertResult = new HashMap<T,TreeSet<T>> ();	

  public TreeSet<T> insert (T e){
    return insertResult.get(e); }
  public void add_insert (T e, TreeSet<T> concVal){
      insertResult.put(e, concVal); }
}
\end{lstlisting} 

Line 19 declares and initializes the attribute that will store the
information about the results of method \javatext{insert} -- for each
value of the parameter \javatext{T e}, it will store a ``concrete''
object. Methods \javatext{insert} and \javatext{add_insert}, in lines
21 to 24, allow to retrieve and define, respectively, the result of
\javatext{insert} for every value of the operation's parameter.

\begin{figure*}[h]
\centering
\includegraphics[width=0.8\textwidth]{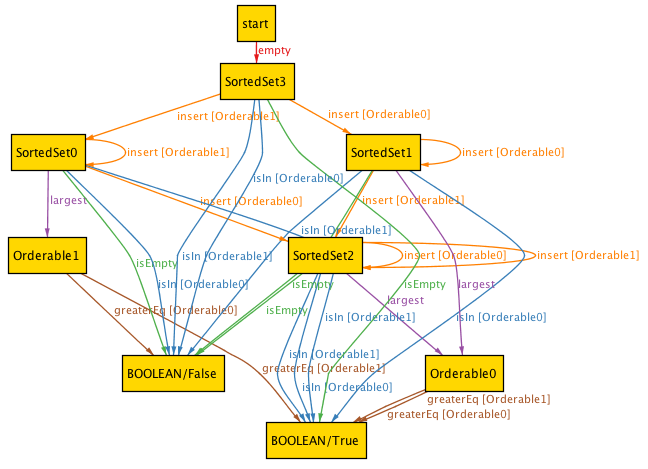}
\caption{A structure satisfying the \congutext{SortedSet} specification module, found by the Alloy Analyzer.}
\label{fig:AlloyModelGeral}
\end{figure*} 

\subsection{Obtaining a collection of instances satisfying the specification}
\label{sec:whatif}

In order to obtain a collection of instances of the specification sorts that conform to our
specification, Flasji capitalizes on the Alloy Analyzer
tool~\cite{Jackson2012} which, if such a finite collection exists, is capable
of generating it (such a finite collection does not exist e.g. in the
case of a specification of an unbounded stack). 

The results presented in this paper assume Flasji asks the
Alloy Analyzer to generate a structure with a fixed number of objects
of the core type; work is under way to optimize the determination of
the number of objects that should be considered of each type.

Figure~\ref{fig:AlloyModelGeral} shows an example for the
\congutext{SortedSet} specification. This collection consists of 2
\congutext{Orderable} instances and 4 \congutext{SortedSet} ones. The
result of every applicable operation is defined for each of these
instances (e.g. \congutext{SortedSet2} contains the two
\congutext{Orderable}s, and it is obtained by inserting \congutext{Orderable0} into \congutext{SortedSet0} or else from inserting \congutext{Orderable1} into \congutext{SortedSet1}).

These instances define correct, expected behaviour; in the next steps,
Flasji will use
the non-core sort ones (\congutext{Orderable} instances
in the example) to create mock ``abstract'' objects that will
represent well-behaved objects, and uses the core sort ones
(\congutext{SortedSet} instances in the example) to create mock
``abstract'' objects (\javatext{TreeSetMock} and ``concrete''
corresponding ones (\javatext{TreeSet} instances
in the example) instances in the example)
that will be compared. Let us see how.

\subsection{Creating the test class}
\label{sec:strategy4}
Flasji generates a test class containing instructions to:
\begin{enumerate}
\item create ``abstract'' objects corresponding to the objects composing the Alloy structure that conforms to the specification;
\item create ``concrete'' objects of the core type
that correspond to the ``abstract'' ones (by using the corresponding
concrete constructors); and
\item compare the behaviour of these
``abstract'' and ``concrete'' objects, by observing them in equal
circumstances, that is, by applying corresponding methods and
comparing the results.
\end{enumerate}

By compiling and executing this test class, Flasji will be able to get
information that it will interpret in order to find the faulty method,
as explained ahead. First let us see how Flasji acomplishes this test
class creation task.

\subsubsection{Creating abstract objects}
\label{sec:strategy41}
Listing~\ref{fig:AbsVSconcTest0} shows part of the generated test
class, namely the creation of the ``abstract'' objects according to the
Alloy structure defined in figure~\ref{fig:AlloyModelGeral}: two
\javatext{OrderableMock} instances (lines 4 and 5) and four
\javatext{TreeSetMock} ones  (lines 11 to 14). 

Lines 6 to 9 show
\javatext{OrderableMock} objects being initialized -- because the only
method of this type is \javatext{greaterEq}, only the method
\javatext{add_greaterEq} is invoked over each ``abstract'' object, for
every possible value of its parameter, in
order to give these objects the information about the expected,
correct, results.

We postpone the initialization of the \javatext{TreeSetMock} objects because it implies previous creation of the corresponding concrete objects.

\begin{lstlisting}[
language=Java, 
breaklines=true,
xleftmargin=5pt,
xrightmargin=0pt,
numbers=left,
numberstyle=\tiny,  
numbersep=5pt,     
captionpos=b, 
caption=(Part of) the test class -- building the ``abstract'' objects (incomplete).,
label=fig:AbsVSconcTest0]
@Test
public void abstractVSconcreteTest () {
  //IOrderable Mocks
  OrderableMock orderable0 = new OrderableMock();
  OrderableMock orderable1 = new OrderableMock();
  orderable0.add_greaterEq(orderable0, true);
  orderable0.add_greaterEq(orderable1, true);
  orderable1.add_greaterEq(orderable0, false);
  orderable1.add_greaterEq(orderable1, true);
  //Abstract objects TreeSet
  TreeSetMock <OrderableMock> sortedSet3 = new TreeSetMock <OrderableMock>();
  TreeSetMock <OrderableMock> sortedSet0 = new ...;
  TreeSetMock <OrderableMock> sortedSet1 = new ...;
  TreeSetMock <OrderableMock> sortedSet2 = new ...;
\end{lstlisting}

\subsubsection{Creating concrete objects}

Flasji also builds ``concrete'' objects for the core sort. For each
``abstract'' object of the core sort there will exist a corresponding
``concrete'' one, which will be built
using the corresponding
constructor methods (see listing~\ref{fig:AbsVSconcTest01}). 

For example, according to the structure in
figure~\ref{fig:AlloyModelGeral}, the \congutext{sortedSet0} instance
of sort \congutext{SortedSet} can be obtained by application of the creator constructor \congutext{empty} followed by application of the transformer constructor \congutext{insert} with parameter \congutext{orderable1}; complying with this (see lines 7 and 8), we build the corresponding ``concrete'' object \javatext{concSortedSet0} using the java constructor \javatext{TreeSet<OrderableMock>()}, which corresponds to the creator constructor \congutext{empty}, and apply to it the method \javatext{insert}, which corresponds to the transformer constructor \congutext{insert}, with parameter \javatext{orderable1}. Whenever there are several ways to build an object, the shortest path is chosen.

\begin{lstlisting}[
language=Java, 
breaklines=true,
xleftmargin=5pt,
xrightmargin=0pt,
numbers=left,
numberstyle=\tiny,  
numbersep=5pt,     
captionpos=b, 
caption=Continuing... building the ``concrete'' objects (incomplete).,
label=fig:AbsVSconcTest01]
@Test
public void abstractVSconcreteTest () {
  ...
  //Create concrete objects TreeSet
  TreeSet<OrderableMock> concSortedSet0 = new TreeSet <OrderableMock>();
  concSortedSet0.insert(orderable1);
  TreeSet<OrderableMock> concSortedSet0_1 = new ...;
  concSortedSet0_1.insert(orderable1);
  ...
  TreeSet<OrderableMock> concSortedSet0_5 = new ...;
  concSortedSet0_5.insert(orderable1);
  // three more to go (concSortedSet3, 1 and 2)...
\end{lstlisting}

Notice that, since methods will be applied to these ``concrete'' objects in order to verify their behaviour, as many copies of a given ``concrete'' object are created as methods applied to it, in order to cope with undesired side effects. 

\subsubsection{Back to abstract objects}
Now that ``concrete'' objects are already created, we can initialize the \javatext{TreeSetMock} objects:

\begin{lstlisting}[
language=Java, 
breaklines=true,
xleftmargin=5pt,
xrightmargin=0pt,
numbers=left,
numberstyle=\tiny,  
numbersep=5pt,     
captionpos=b, 
caption=Continuing... initializing \javatext{TreeSetMock} objects.,
label=fig:AbsVSconcTest2]
@Test
public void abstractVSconcreteTest () {
  ...
  //Initializing sortedSet abstract objects
  sortedSet0.add_isEmpty(false);
  sortedSet0.add_largest(orderable1);
  sortedSet0.add_isIn(orderable0, false);
  sortedSet0.add_isIn(orderable1, true);
  sortedSet0.add_insert(orderable0, concSortedSet2);
  sortedSet0.add_insert(orderable1, concSortedSet0);
// three more to go (sortedSet3, 1 and 2)...
\end{lstlisting}

Lines 5 to 8 ``feed'' the \javatext{sortedSet0} object with
the information that it represents a sorted set that is not empty, whose
largest element is the \javatext{orderable1} object, and that it
contains \javatext{orderable1} but not \javatext{orderable0}, as
would be expected by inspection of the structure
in figure~\ref{fig:AlloyModelGeral}.  In the case of object
\javatext{sortedSet3}, which is empty as can be seen in
figure~\ref{fig:AlloyModelGeral}, the instruction invoking the method
\javatext{add_largest()} over it would not be generated, since the
operation \javatext{largest} is undefined for that object.

These
informations 
will be used later on to obtain the values that are expected to be the results
of the corresponding methods when applied to the ``concrete'' object
corresponding to \javatext{sortedSet0}, which, by construction, is
\javatext{concSortedSet0} (or any of its copies).

Line 9 ``feeds'' \javatext{sortedSet0} with the information
about which ``concrete'' object should be expected after inserting
\javatext{orderable0} in the ``concrete'' object that corresponds to
\javatext{sortedSet0} (\javatext{concSortedSet0} or any of
its copies) -- the expected result is \javatext{concSortedSet2}. 
In the same way, in line 10, the expected result of inserting \javatext{orderable1} in the ``concrete'' object that corresponds to \javatext{sortedSet0} is defined to be itself.

\subsubsection{Comparing abstract and concrete objects}
In a next step, Flasji generates instructions in the test class that
invoke all possible operations over the ``abstract'' and ``concrete''
objects and compare the results:

\begin{lstlisting}[
language=Java, 
breaklines=true,
xleftmargin=5pt,
xrightmargin=0pt,
numbers=left,
numberstyle=\tiny,  
numbersep=5pt,     
captionpos=b, 
caption=Continuing... comparing ``abstract'' and ``concrete'' objects (incomplete).,
label=fig:AbsVSconcTest1]
@Test
public void abstractVSconcreteTest () {
  ...
  //Compare concrete with corresponding abstract
  assertTrue(concSortedSet0_1.isEmpty() == sortedSet0.isEmpty());
  assertTrue(concSortedSet0_2.largest() == sortedSet0.largest());
  assertTrue(concSortedSet0_3.isIn(orderable1) == sortedSet0.isIn(orderable1));
  assertTrue(concSortedSet0_4.isIn(orderable0) == sortedSet0.isIn(orderable0));
  concSortedSet0_5.insert (orderable1);
  assertTrue(concSortedSet0_5.equals(sortedSet0.insert (orderable1)));  
  //three more to go (concSortedSet3, 1 and 2)...
\end{lstlisting}
 
The JUnit method \javatext{assertTrue} is used to generate an
\javatext{AssertionError} exception
whenever the behaviour of the ``concrete'' objects is not as expected,
that is, whenever the results of methods invoked over ``concrete''
objects are different from the ones indicated by their ``abstract''
counterparts.

Lines 5 to 8 show the comparison between the \javatext{sortedSet0}
``abstract'' object and its ``concrete'' counterpart \javatext{concSortedSet0} 
using each \javatext{TreeSet} method
whose result type is not \javatext{TreeSet} nor \javatext{void}. Since
all these results are of primitive types or of the parameter type
\javatext{OrderableMock}, Flasji uses \javatext{==} to compare between
\javatext{sortedSet0} and \javatext{concSortedSet0} results. 

Lines 9 and 10 show the comparison between ``abstract'' and ``concrete'' objects using (the only) operation with a core result type -- \javatext{insert}.
As already referred, to verify whether a given operation whose result is of the core type is well implemented, we compare the ``concrete'' object the method returns, with the ``concrete'' object that it should return. Since \javatext{insert} is implemented with a \javatext{void} result type, we must first invoke the method using the ``concrete'' object as a target, and
then we compare (using \javatext{equals}) its new state with the ``concrete'' object that, according to the ``abstract'' corresponding object, should be the correct result.

Since the ultimate goal of this test class is to find the method
containing the fault, it should be possible to reason about the
results of all these comparisons, so we must be able to test all the
\javatext{assert} commands. Although we do not show it in this paper
due to space limitations, enclosing each \javatext{assertTrue}
invocation in a \javatext{try-catch} block that catches
\javatext{AssertionError} exceptions, allows to collect all results
which will help composing a final test diagnosis.

A final note before continuing: whenever the module of Congu
specifications input to Flasji includes more than one non-parameter
type, e.g., the case where the input includes a core sort \congutext{C} and one non-core, non-parameter sort \congutext{N},
the class implementing \congutext{C} is verified for faults considering that the class
implementing \congutext{N} is correct. No mock class is built for \congutext{N}, hence no
``abstract'' \javatext{N} objects are created; only ``concrete''
\javatext{N} objects are. For methods that return \javatext{N} type results,
``abstract'' \javatext{C} objects  are ``fed'' with the information about which \javatext{N}
``concrete'' object should be the expected result. Thus, comparison
between actual and expected results relating these methods are achieved
using \javatext{equals}. The running \congutext{SortedSet} example does not cover this kind
of situation.

\subsection{Running the test and interpreting the results}
\label{sec:results}

As soon as the test class is generated, Flasji compiles it and
executes it. Then, it interprets the results of the tests. The
interpretation is based upon the following
observations:
\begin{enumerate}
\item whether several and varied observers (non-constructor operations) fail or only
one fails -- this is important to decide whether to blame a
constructor or a given, specific, observer;
\item whether varied
observers fail when applied to ``concrete'' objects created only by
the constructor-creator, or when applied to objects that were also the
target of non-creator constructors -- this is important to decide
which constructor is the faulty one.
\end{enumerate}

The result interpretation algorithm inspects
three data structures containing data collected during the execution of
the test (whenever an \javatext{assertTrue} command fails):

\begin{itemize}
\item $L_1$ - Set of pairs $< obs ; obj >$ that register that
  differences occurred between expected and actual behaviour, for given observer $obs$ and object
  $obj$;
\item $L_2$ - Set of pairs $< ncc ; obj >$ that register that
  differences occurred between expected and actual
  behaviour, for given non-creator (transformer) constructor $ncc$ and object
  $obj$;
\item $L_3$ - Set of pairs $< cc ; n >$ that register for every
  creator constructor $cc$ the number of failed observations over objects uniquely
  built with $cc$;
\end{itemize}

If, when applied to concrete objects, more than one observer methods
present results that are different from the ones expected ($(L_1
\cup L_2)$ contains pairs for more than one observer), we may
infer that the method(s) used to build those concrete objects are ill-implemented, and that the problem does not come from some particular way of inspecting the objects.
If the implementation of a given observer is wrong, one would not
expect problems when inspecting the objects using the other observers,
but only in the observations involving that particular one.

If a constructor-creator $cc$ (in the running example,
 \javatext{TreeSet()} is the $cc$ that implements the \congutext{empty} creator
 operation) is faulty, it is reasonable to think that the application
 of the other constructors over an object created with $cc$ will most
 probably result in non-conformant objects, because the initial object
 is already ill-built. The information in $L_3$ allows us to focus on creator-constructors.

When no problems arise when observing a freshly created object, but
they do arise when observing those objects after being affected by a given non-creator constructor $ncc$  (\javatext{insert} in the running example), then one may point the finger to $ncc$.

\begin{tabbing}
\ \ \ \  \= \ \ \ \   \=  \ \ \ \  \= \ \ \ \  \= \\
\textbf{if} ($L_1 \cup L_2$) contains pairs for more than 1 observer, \textbf{then}\>  \>  \>  \> \\
\> \textbf{if} there exists $<cc,i>$ in $L_3$ with $i>0$, \textbf{then}  \>  \>  \> \\
\>  \>  \textbf{if} that pair $<cc,i>$ with $i>0$ is unique, \textbf{then} \>  \> \\
\>  \>  \> $cc$ is guilty; \> \\
\>  \> \textbf{else} \> \> \\
\>  \>  \> inconclusive; \> \\
\>  \> \textbf{endIf} \> \> \\
\> \textbf{else} \> \> \> \\
\> \> \textbf{for} each non-creator constructor $ncc_j$ \textbf{do} \>
\> \\
\>  \>  \> $L_{nccj} \leftarrow$ sub-set of $L_2$ containing only
pairs from $L_2$ whose first  element is $ncc_j$;\> \\
\>  \>  \> Delete from $L_{nccj}$ the pairs whose $obj$ was not built using only $ncc_j$ and a creator constructor;\> \\
\>  \>  \> \textbf{if} $L_{nccj}$ is not empty, \textbf{then} \> \\
\>  \>  \> \> add $ncc_j$ to the final set of suspects (FSS);\\
\>  \> \> \textbf{endIf} \> \\
\>  \> \textbf{endFor} \> \> \\
\> \textbf{endIf} \> \> \> \\
\> \textbf{if} $\#$FSS = 1 \textbf{then} \> \> \> \\
\>  \> the guilty is the sole element of FSS; \> \> \\
\> \textbf{else} \> \> \> \\
\>  \> inconclusive; \> \> \\
\> \textbf{endIf} \> \> \> \\
\textbf{else} \> \> \> \> \\
\> \textbf{if} ($L_1 \cup L_2$) is empty, \textbf{then} \> \> \> \\
\>  \> inconclusive; \> \> \\
\> \textbf{else} \> \> \> \\
\>  \> the guilty is the sole observer in ($L_1 \cup L_2$); \> \> \\
\> \textbf{endIf} \> \> \> \\
\textbf{endIf} \> \> \> \> \\
\end{tabbing}

\vspace{-15pt}  
If the algorithm elects a guilty method in the end, then the
user is given the identified method as the most
probable guilty. 
In either case, the set FSS of (other) suspects is
presented.


\section{Evaluation}
\label{sec:evaluation}
To evaluate the effectiveness of our approach, we applied it to two case studies -- this paper's \congutext{SortedSet} running example, and a \congutext{MapChain} specification module and corresponding implementations. The Java classes implementing the designated sorts of both case studies where seeded with faults covering all the specification operations. 

We put Flasji to run for every defective class, and registered the outputs.

We also tested those defective classes in the context of two existing fault-location tools -- GZoltar~\cite{Abreu2009,RiboiraA10} and EzUnit4~\cite{Bouillon2007,Steimann2009} --, that give as output a list of methods suspect of containing the fault, ranked by probability of being faulty. The tests suites we used were generated by the GenT~\cite{Andrade2012,Andrade2011} tool (already referred to in this paper), from the ConGu specifications and refinement mappings; under given restrictions (e.g., the specification has finite models) GenT generates comprehensive test suites, that cover all specification axioms. GenT generated 20 test cases for the \congutext{SortedSet} case, and 17 for the \congutext{MapChain} one.

Finally we compared the three tools' results for every defective variation of each case study.

For each of the defective versions of the designated sorts implementations (for example, two different faults were seeded in \congutext{SortedSet} \javatext{isEmpty} method, three in \congutext{MapChain} \javatext{get} method, etc) table~\ref{tab:comparative} shows:
\begin{itemize}
\item the number of tests (among the 20 JUnit tests that were generated by GenT for the \congutext{SortedSet} case, and 17 for the \congutext{MapChain} one) that failed when both GZoltar and EzUnit4 run them;
\item whether the faulty method was ranked, by each tool, as most probable guilty ($1_{st}$), second most probable guilty ($2_{nd}$) or third or less probable ($n_{th}$). A fourth type of result -- ``No'' -- means the guilty method has not been ranked as suspect at all.
\end{itemize}

\begin{table} [h]
  \centering
\small

  \begin{tabular}{|c|l|c|ccccccccc|} \hline
   & \multicolumn{1}{c|}{\multirow{2}{*}{\textbf{ }}}
    & \multicolumn{1}{c|}{\multirow{2}{*}{\textbf{ }}} 
    & \multicolumn{3}{c|}{\textbf{Faulty method ranked:}} \\ \cline{4-6}
    &\multicolumn{1}{c|}{\textbf{\hspace{8 pt}Faulty method~\hspace{8 pt}}}
    & \multicolumn{1}{c|}{\textbf{\hspace{4 pt}failed tests~\hspace{4 pt}}} 
    & \multicolumn{1}{c|}{\textbf{\hspace{8 pt}Flasji~\hspace{8 pt}}} 
    & \multicolumn{1}{c|}{\textbf{\hspace{7 pt}EzUnit4~\hspace{7 pt}}} 
    & \multicolumn{1}{c|}{\textbf{\hspace{6 pt}GZoltar~\hspace{6 pt}}}
     \\ \hline

     \multirow{7}{*}{\begin{sideways} \congutext{SortedSet}\end{sideways}} 
    & \hspace{4 pt} isEmpty & 5 & $1_{st}$ & $n_{th}$ & \multicolumn{1}{c|}{$2_{nd}$} \\
    & \hspace{4 pt} isEmpty & 1 & $1_{st}$ & $1_{st}$ & \multicolumn{1}{c|}{$2_{nd}$} \\
    & \hspace{4 pt} isIn & 1 & $1_{st}$ & $n_{th}$ & \multicolumn{1}{c|}{$1_{st}$} \\
    & \hspace{4 pt} largest & 7 & $1_{st}$ & $1_{st}$  & \multicolumn{1}{c|}{$1_{st}$} \\
    & \hspace{4 pt} largest & 1 & $1_{st}$ & $1_{st}$  & \multicolumn{1}{c|}{$2_{nd}$} \\
    & \hspace{4 pt}\textit{private} insert & 2 & No & $n_{th}$ & \multicolumn{1}{c|}{$2_{nd}$} \\
    & \hspace{4 pt}\textit{public} insert & 5 & $1_{st}$ & $n_{th}$ & \multicolumn{1}{c|}{$2_{nd}$} \\ \hline
    \multirow{10}{*}{\begin{sideways} \congutext{MapChain}\end{sideways}} 
    & \hspace{4 pt} get & 4 & No & $n_{th}$ & \multicolumn{1}{c|}{$1_{st}$} \\
    & \hspace{4 pt} get & 3 & No & $2_{nd}$ & \multicolumn{1}{c|}{$1_{st}$} \\
    & \hspace{4 pt} get & 4 & No & $n_{th}$ & \multicolumn{1}{c|}{$n_{th}$} \\
    & \hspace{4 pt} isEmpty & 2 & $1_{st}$ & $2_{nd}$ & \multicolumn{1}{c|}{$1_{st}$} \\
    & \hspace{4 pt} isEmpty & 1 & $1_{st}$ & $n_{th}$ & \multicolumn{1}{c|}{$1_{st}$} \\
    & \hspace{4 pt} put & 0 & $1_{st}$ & No & \multicolumn{1}{c|}{No} \\
    & \hspace{4 pt} put & 1 & $1_{st}$ & $1_{st}$ & \multicolumn{1}{c|}{$2_{nd}$} \\
    & \hspace{4 pt} put & 2 & $1_{st}$ & $1_{st}$ & \multicolumn{1}{c|}{$2_{nd}$} \\
    & \hspace{4 pt} remove & 1 & $1_{st}$ & $2_{nd}$ & \multicolumn{1}{c|}{$2_{nd}$} \\
    & \hspace{4 pt} remove & 1 & $1_{st}$ & $2_{nd}$ &\multicolumn{1}{c|}{$2_{nd}$} \\ \hline
  \end{tabular}

  \caption{Results of comparative experiments. ``$1_{st}$'', ``$2_{nd}$'' and ``$n_{th}$'' stand for first, second and third or worse, respectively. ``No'' means the faulty method has not been ranked as suspect.}
  \label{tab:comparative}
\end{table}

\begin{flushleft}
\begin{figure}[h]
\centering
\includegraphics[width=0.8\textwidth]{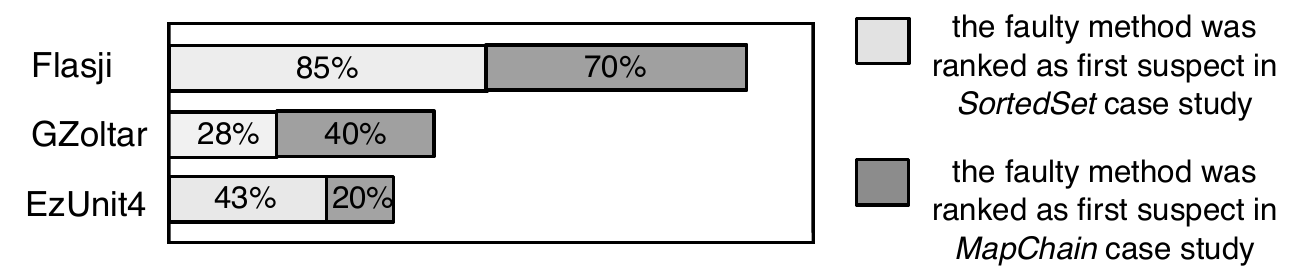}
\caption{Summary of the evaluation experiment. The bars measure the success of each approach in ranking the faulty method as first suspect.}
\label{fig:FljVSothers}
\end{figure} 
 \end{flushleft}

Flasji provided very accurate results in general (see also a summary
in figure~\ref{fig:FljVSothers}). The bad results in the three faults
for method \javatext{get} of the \congutext{MapChain} case study
(there were no suspects found whatsoever) are due to the fact that
\javatext{equals} uses the \javatext{get} method, therefore becoming
unreliable whenever method \javatext{get} is faulty. This case
exemplifies the \textit{oracle problem} (see
section~\ref{sec:rw}). 

Applying an alternative method of observation
(see~\cite{Nunes2012}) -- one where the \javatext{equals} method is
not used and, instead, only the outcomes of observers whose result is
not of the core sort are used in comparisons -- we obtain the right
results for this case, i.e. \javatext{get} is ranked as prime
suspect. However, the good results we had for the 3rd faulty
\javatext{put} method and the 1st \javatext{remove} got worse -- they
are ranked second instead of first. These particular cases indicated
\javatext{isEmpty} as prime suspect because the seeded fault of both
those methods was the absence of change in the number of elements in
the map whenever insertion/removal happens, which made
\javatext{isEmpty} fail.

Another critical issue w.r.t. our approach is the one concerned with private methods. The fault in \emph{private} method \javatext{insert} of the \congutext{SortedSet} case study, caused Flasji to rank the \emph{public} \javatext{insert} method, instead, as the most probable suspect (in the particular implementation used, the public \javatext{insert} method is composed of one only statement which invokes the private \javatext{insert} method). As expected, private methods are not identified as suspects by Flasji because they do not directly refine any specification operation (as defined in the refinement mapping from specifications to implementations); instead, the public, specified, methods that invoke them are identified.
 
A case worth mentioning is the one corresponding to the first seeded fault in the \javatext{put} method of the \congutext{MapChain} case study, where none of the seventeen GenT tests fail (the particular case that causes the error was not covered). As a consequence, neither EzUnit4 and GZoltar detected the fault; on the contrary, Flasji succeeded in detecting the guilty method.

\section{Related work}
\label{sec:rw}

The approach presented in this paper relies on the existence of structures satisfying the specification to supply the behaviour of objects to be used in tests. The structured nature of specifications, where functions and axioms are defined sort by sort, and where the latter are independently implemented by given Java types, is essential to the incremental integration style of Flasji. 

Several approaches to testing implementations of algebraic
specifications exist, that cover test generation (\cite{Claessen2000,Doong1994,Gaudel2008,Hughes1996,Kong2007} to name a few), and many compare the two sides of equations where variables have been substituted by ground terms -- differences exist in the way ground terms are generated, and in the way comparisons are made. The gap between algebraic specifications and implementations makes the comparison between concrete objects difficult, giving rise to what is known as the \textit{oracle problem}, more specifically, the search for reliable decision procedures to compare results computed by the implementation. Whenever one cannot rely on the \javatext{equals} method, there should be another way to investigate equality between concrete objects. Several works have been proposed that deal with this problem, e.g.~\cite{Gaudel2008,Machado2002,Zhu2003}. 
In~\cite{Nunes2012} we tackle this issue by presenting an alternative
way of comparing concrete objects, one that relies only in observers
whose result is of a non-core sort. In some way this complies with the
notion of observable contexts in~\cite{Gaudel2008} -- all observers
but the ones whose result is of the designated sort constitute
observable contexts.

The unreliability of \javatext{equals} can also affect the
effectiveness of the GenT tests~\cite{Andrade2012} since this method is used whenever
concrete objects of the same type are compared. 
One of the
improvements we intend to make is to give Flasji the ability to test the \javatext{equals} method in order to make its use more reliable. 

\section{Conclusions}
\label{sec:conclusoes}

We presented Flasji, a technique whose goal is to test Java
implementations of algebraic specifications and find the method that
is responsible for some deviation of the expected behaviour.  

Flasji capitalyzes on ConGu, namely using ConGu specification and
refinement languages, and enriches it with the capability of finding
faulty methods. It accomplishes the task through the generation of
tests that are based on structures satisfying the specification. The behaviour of instances of the
implementation is compared with the one expected, as given by those
specification-compliant structures. 
The results of the comparisons are interpreted in order to
find the method responsible for the fault.

An evaluation experiment was
presented where Flasji results over two case studies, for which faults have been
seeded in the implementing Java classes, are compared with two other
tools' results when executed over comprehensive suites of tests.
The encouraging results obtained in comparative studies led us to
continue working on it, with the purpose of improving some negative aspects and weaknesses, some of which already identified and reported
in this paper. 

The following improvements, among others, are planned:
(i) testing the implementation of the \javatext{equals}
method, even if the specification module does not specify it, in order
to be able to better rely on its results, (ii) optimizing the
determination of the number of objects of each type that an Alloy
structure conforming to the specification
should contain (the results here presented assumed Flasji asks the
Alloy Analyzer to generate a structure with a fixed number of objects
of the core type), and (iii) whenever there are several non-parameter
types, apply the process several times, each considering one of them
as the core type, and integrate the results (special cases as
e.g. inter-dependent types, deserve attention).

\bibliographystyle{eptcs}
\bibliography{faultLocation}
\end{document}

%% file: macros.tex
\usepackage{xspace}
\usepackage{graphicx}

\lstdefinelanguage{Congu}{
  extendedchars=true,
  showstringspaces=false,
  basicstyle=\sffamily\upshape\scriptsize,
  morekeywords={specification,sorts,constructors,observers,others,axioms,domains,end,not,and,or,if,iff,else,when,abs},
  morekeywords={refinement,is,import, extends},
  morekeywords={module,core,parameter},
  literate={-->}{{$\longrightarrow$}}3,
  morecomment=[l]{//},
  morecomment=[s]{/*}{*/}, 
}

\lstdefinelanguage{Alloy}{
  extendedchars=true,
  showstringspaces=false,
  basicstyle=\sffamily\upshape\scriptsize,
  morekeywords={open,one,lone,sig,fact,in,extends,Int,no,all,implies,iff,else,and,or,run,for,but,exactly,some},
  morecomment=[l]{//},
  morecomment=[s]{/*}{*/}, 
}

\newcommand{\congutext}[1]{\lstinline[language=Congu,
  basicstyle=\sffamily\upshape\small]|#1|}

\newcommand{\javatext}[1]{\lstinline[language=Java,
  basicstyle=\ttfamily\upshape\small]|#1|}

